\documentclass[graybox]{svmult}
\usepackage{helvet}   
\usepackage{courier}            
\usepackage{type1cm}  

\usepackage{amsmath,amssymb,url}
\newbox\shell
\newcommand{\hf}{$\frac12$}
\newcommand{\mrk}[3]{\put(#1,#2){\mbox{#3}}}
\newcommand{\dia}[2]{\setbox\shell=\hbox{\begin{picture}(180,220)(-90,-110)#1
\put(-90,-95){\makebox(180,220)[b]{$#2$}}\end{picture}}\dimen0=\ht
\shell\multiply\dimen0by7\divide\dimen0by16\raise-\dimen0\box\shell\hfill}
\newcommand{\tet}{\put(-100,-50){\line(1,0){200}}
\put(-100,-50){\line(2,3){100}}\put(100,-50){\line(-2,3){100}}
\put(-100,-50){\line(2,1){100}}\put(100,-50){\line(-2,1){100}}
\put(0,0){\line(0,1){100}}}
\newcommand{\tetmrk}[6]{\tet
\mrk{-75}{24}{#1}\mrk{-55}{-10}{#2}\mrk{2}{24}{#3}
\mrk{57}{24}{#4}\mrk{45}{-10}{#5}\mrk{-10}{-80}{#6}}

\begin{document}

\title*{Multivariate elliptic kites and tetrahedral tadpoles}
\author{David Broadhurst}
\institute{David Broadhurst \at Open University, Milton Keynes MK76AA, UK, \\ 
\email{David.Broadhurst@open.ac.uk}}

\maketitle

\abstract{This work deals with two types of Feynman integrals in perturbative quantum field theory: 
the 2-loop 2-point kite, with 5 arbitrary internal masses, and its completion by a sixth propagator, 
to give a 3-loop tetrahedral tadpole, with 6 arbitrary masses.
These general-mass cases cover broken and
unbroken gauge theories,  based on the
Lie algebras $U(1)$, $SU(2)$ and $SU(3)$, for the electromagnetic, weak
and strong interactions.
The elliptic substructure of these integrals should not be regarded as an obstruction.
Rather, it is a bonus, thanks to the arithmetic-geometric mean of Gauss.
Compact formul{\ae} are given, to handle all cases. Zero-mass limits are carefully considered.
Anomalous thresholds of triangles in the kite pose no problem. 
The number theory of tadpoles is investigated, with intriguing results.}

\section{Introduction}

A univariate elliptic kite in the two-loop electron propagator
was tackled by Afaf Sabry~\cite{S} sixty years ago, after work with Gunnar K\"all\'en~\cite{KS} on the two-loop photon propagator.
The latter has an evaluation in terms of the classical trilogarithm of Leonard Lewin's influential books~\cite{L}. Thirty years later~\cite{B90},
I extended both results to include all cases needed in unbroken gauge theories.
Dispersion relations, pioneered by Hans Kramers and Ralph Kronig~\cite{KK}
and extended to particle physics by Murph Goldberger~\cite{MG}, were of the essence in this work. I learnt about them from
Gabriel Barton~\cite{GB} who had keenly attended Goldberger's lectures in Princeton.

Work on univariate $L$-loop equal-mass sunrise integrals~\cite{SR1,SR2}, with $L+1$ edges,
was a vital part of the preparation for Stefano Laporta's result for the magnetic moment of the electron at 4 loops~\cite{SL}.
This subject was revolutionized by Spencer Bloch and Pierre Vanhove~\cite{BV}, who gave a modular parametrization
of the 2-loop sunrise diagram in two spacetime dimensions.  With Matt Kerr, they achieved the same feat
at 3 loops~\cite{BKV} where the underlying differential equation is a symmetric square of the equation at 2 loops, as had been
observed by Geoff Joyce~\cite{GJ}, long before, in the context of lattice Green functions for condensed matter problems.
With David Bailey, Jon Borwein and Larry Glasser~\cite{BBBG} I exploited this in elliptic evaluations of Bessel moments,
which are related to Feynman integrals by the fact that the Fourier transform of a massive propagator
is a Bessel function. Many conjectures~\cite{Bc3,Bc2,Bc4,Bc1} on Bessel moments were later proved by Yajun Zhou~\cite{YZ1,YZ2,YZ3,YZ4}.

Rich mathematical structure also resides in the 6-parameter tetrahedron made by completing the kite with a sixth
propagator, to obtain a 3-loop vacuum integral which is called a tadpole. In~\cite{B92,B99,B93}, I considered binary
tetrahedral tadpoles, with unit or zero masses, discovering that all are reducible to multiple polylogarithms
of sixth roots of unity, despite elliptic substructure in constituent kites.
These results were used by Matthias Steinhauser, whose computer-algebra package~\cite{MS}  
reduces every binary 3-loop tadpole, no matter what its subdivergences, to 
polylogarithmic constants identified in~\cite{B99}.

Activity on elliptic substructure of massive Feynman integrals overlaps with work in perturbative string theory~\cite{St2,St1,St3},
on massless Feynman integrals, where elliptic substructure appears in a 2-loop 6-point function with 4 off-shell legs~\cite{SYM2,SYM1}, 
and on developing~\cite{Dev4,Dev3,Dev1,Dev5,Dev2} a theory of iterated integrals with elliptic kernels, 
as generalizations of better understood multiple polylogarithms~\cite{MPL4,MPL2,SC,MPL5,MPL3,MPL1}. 

Participation at meetings organized in Berlin, Bonn, Copenhagen, Crieff,
Ettal, Florence, Geneva, Grenoble, Mainz, M\"unster, Padua, Potsdam, Siegen, Varna, Zeuthen and Z\"urich revealed
the breadth and depth of activity in these related fields. 
For multiloop sunrise diagrams~\cite{Sm1,Sm2,Sm3}, with no internal vertices, multivariate
analysis is not so hard. For the generic tetrahedral completion of a kite, each of the
four faces specifies a subintegration with square roots associated to each of the three
vertices that define the face. Yet the net result may still be polylogarithmic, notwithstanding
elliptic substructure.

The remainder of this article is organized as follows. After some short preliminary definitions,
Section~2 gives compact explicit formul{\ae} that enable one to evaluate any massive kite or
tetrahedral tadpole numerically, to high precision and at good speed, by one-dimensional
quadrature of the derivative $\sigma^\prime$ of the discontinuity of a kite, multiplied by
a logarithm or dilogarithm. The elliptic term in the integrand presents no problem,
since it is given by the extraordinarily fast process of the arithmetic-geometric mean of Gauss~\cite{AGM1,AGM2}.  
Nor are the anomalous thresholds of triangles problematic.
Section~3 gives zero-mass limits needed for gauge theories.
Section~4 is devoted to the number theory of tetrahedral tadpoles. Section~5 gives
comments and a summary.

\subsection{Preliminaries}

I define the two-loop scalar kite integral in 4-dimensional Minkowski space
\begin{gather}I(q^2;m_1^2,m_2^2,m_3^2,m_4^2,m_5^2)=-\frac{q^2}{\pi^4}\int{\rm d}^4l\int{\rm d}^4k\,
\prod_{j=1}^5\frac{1}{p_j^2-m_j^2-{\rm i}\epsilon},\label{idef}\\
(p_1,\,p_2,\,p_3,\,p_4,\,p_5)=(l,\,l-q,\,l-k,\,k,\,k-q),\label{pj}
\end{gather}
with a labelling chosen such that edge 1 is not adjacent to edge 5. Neither is edge 2 adjacent to edge 4.
Where appropriate, the arguments $m_j^2$ will be omitted.
Then $I(s)$ is analytic in the $s$-plane with a cut $s\in[s_{\rm L},\infty]$ and a leftmost branch point $s_{\rm L}$ that
is the lowest of the thresholds $\{s_{1,2},s_{4,5},s_{2,3,4},s_{1,3,5}\}$, where $s_{j,k}=(m_j+m_k)^2$
and $s_{i,j,k}=(m_i+m_j+m_k)^2$. 

On the top lip of the cut, let $\Im I(s+{\rm i}\epsilon)=\pi\sigma(s)$. In the general mass case, with $\sigma(s_{\rm L})=0$,
it suffices to know the derivative $\sigma^\prime(s)$, which gives the dispersion relation
\begin{equation}I(q^2)=-\int_{s_{\rm L}}^\infty{\rm d}s\,\sigma^\prime(s)\log\left(1-\frac{q^2}{s}\right).
\label{iq}\end{equation}
\mbox{\hspace{1cm}}\hfill
\dia{\tetmrk123456}{}
\dia{\tetmrk213546}{}
\dia{\tetmrk345261}{}
\mbox{\hspace{1cm}}
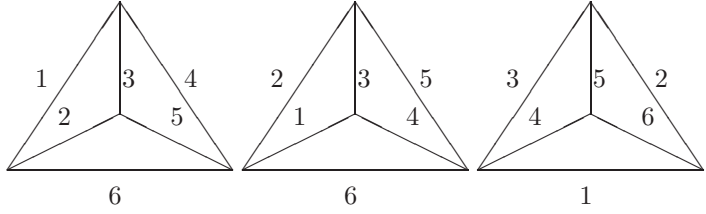
\begin{center}{\bf Figure 1:} Equivalent labellings of a tetrahedral tadpole\end{center}

Consider the logarithmically divergent tetrahedral tadpole formed by joining the external vertices of the kite with
a propagator $1/(q^2-m_6^2)$.
Regularization in $4-2\epsilon$ dimensions gives a tadpole
\begin{gather}
T_{1,2,3}^{5,4,6}=\left(\frac{1}{3\epsilon}+1\right)6\zeta_3+3\zeta_4-F_{1,2,3}^{5,4,6}+O(\epsilon),\label{tad}\\
F_{1,2,3}^{5,4,6}=\int_{s_{\rm L}}^\infty{\rm d}s\,\sigma^\prime(s;m_1^2,m_2^2,m_3^2,m_4^2,m_5^2)
\left({\rm Li}_2\left(1-\frac{m_6^2}{s}\right)+\frac12\log^2\left(\frac{\overline{m}^2}{s}\right)\right)
\label{f123}
\end{gather}
where $\overline{m}$ is the scale of dimensional regularization and the dilogarithm is the analytic
continuation of the sum ${\rm Li}_2(z)=\sum_{n>0}z^n/n^2$, valid for $|z|<1$. The superscripts
in~(\ref{f123}) are made redundant by the labelling convention: a superscript $j$ sits above a subscript $k$
if and only if edge $j$ is not adjacent to edge $k$, with subscripts defining a triangle. 

Figure~1 shows that $F_{1,2,3}=F_{2,1,3}=F_{3,4,5}$. The first identity is obvious from the point of view
of the kite in the integrand of~(\ref{f123}). The second is highly non-trivial, since now $m_1$
is external. The relation $F_{1,2,3}=F_{1,4,6}$ makes $m_3$ external.
Thus a single omnivorous tadople can digest 6 distinct kites, each obstructed by a pair of elliptic curves.

\section{The generic kite}

Consider the generic 5-parameter kite, with distinct non-zero masses and thresholds.
The derivative $\sigma^\prime(s)$ of its discontinuity
receives logarithmic contributions from two-particle cuts and elliptic contributions
from three-particle cuts. There may also be an algebraic
contribution associated with anomalous thresholds in the dispersion relations
for triangles that form the kite. I now give explicit compact formul{\ae} for these three
types of contribution.
 
\subsection{Non-elliptic non-anomalous contribution}

In the absence of anomalous thresholds, the non-elliptic contribution
to $\sigma^\prime(s)$ may be obtained by diligent algebra and application of Cauchy's integral theorem.
Here I give the 5-parameter result, written as compactly as possible.

Denote the square root of the symmetric K\"all\'en function~\cite{Kb} by
\begin{equation}\Delta(a,b,c)=\sqrt{a^2+b^2+c^2-2(ab+bc+ca)}\label{del}\end{equation}
with abbreviations $\Delta_{j,k}(s)=\Delta(s,m_j^2,m_k^2)$ and $\Delta_{i,j,k}=\Delta_{j,k}(m_i^2)$. Then
\begin{equation}D_{j,k}(s)=\frac{r}{s-(m_j-m_k)^2}\log\left(\frac{1+r}{1-r}\right),\quad 
r=\left(\frac{s-(m_j-m_k)^2}{s-(m_j+m_k)^2}\right)^{1/2}\label{f12}\end{equation}
is analytic in the $s$-plane with a cut $s\in[(m_j+m_k)^2,\infty]$, where its discontinuity is
$D_{j,k}(s+{\rm i}\epsilon)-D_{j,k} (s-{\rm i}\epsilon)=-2\pi{\rm i}/\Delta_{j,k}(s)$. Define the real constants
\begin{equation}\alpha=\frac{(m_1^2-m_4^2)(m_2^2-m_5^2)}{m_3^2}-m_3^2,\, 
\beta=\frac{(m_1^2m_5^2-m_2^2m_4^2)(m_1^2-m_2^2-m_4^2+m_5^2)}{m_3^2}
\label{albe}\end{equation}
which help to condense the result. The (possibly complex) constants
\begin{equation}s_\pm=\frac{m_1^2+m_2^2-2m_3^2+m_4^2+m_5^2-\alpha}{2}
\,\pm\,\frac{\Delta_{1,3,4}\Delta_{2,3,5}}{2m_3^2}\label{spm}\end{equation}
locate leading Landau singularities~\cite{AF} of triangles that form the kite.

In the absence of anomalous thresholds, the non-elliptic contribution is 
\begin{equation}\sigma^\prime_{\rm{N}}(s)=
\Theta(s-s_{1,2})\sigma_{1,2}^\prime(s)+
\Theta(s-s_{4,5})\sigma_{4,5}^\prime(s)\label{sn}\end{equation}
with Heaviside steps denoted by  $\Theta$. Putting edges 1 and 2 on-shell, I obtain
\begin{gather}\Delta_{1,2}(s)\sigma_{1,2}^\prime(s)=\Re\left((s+\alpha)D_{4,5}(s)
+L_{4,5}+\sum_{i=0,+,-}C_i\,\frac{D_{4,5}(s)-D_{4,5}(s_i)}{s-s_i}\right),\label{s12}\\
C_\pm=\alpha s_\pm +\beta,\;C_0=(m_2^2-m_1^2)(m_4^2-m_5^2),\;s_0=0,\;
L_{4,5}=\log\left(\frac{m_4m_5}{m_3^2}\right) .\label{c0pm}\end{gather}
To obtain $\sigma_{4,5}^\prime$, exchange $(m_1,m_2)\longleftrightarrow(m_4,m_5)$,
noting that this does not alter the coefficients $C_i$ or the arguments $s_i$.

The result in~(\ref{sn},\ref{s12}) holds if both of the conditions 
\begin{eqnarray}
(m_1+m_2)(m_3^2+m_1m_2)&\ge& m_1m_5^2+m_2m_4^2\label{ok12}\\
(m_4+m_5)(m_3^2+m_4m_5)&\ge& m_5m_1^2+m_4m_2^2\label{ok45}
\end{eqnarray}
are satisfied. However, these are not necessary conditions. The final
decision as to whether the non-elliptic contribution is in need of an anomalous correction
is arbitrated by an elliptic contribution that is unambiguous.

\subsection{Elliptic contribution}

The elliptic contribution, from 3-particle intermediate states, has the form
\begin{equation}\sigma^\prime_{\rm{E}}(s)=
\Theta(s-s_{2,3,4})\sigma_{2,3,4}^\prime(s)+
\Theta(s-s_{1,3,5})\sigma_{1,3,5}^\prime(s).\label{sige}\end{equation}
It contains complete elliptic integrals of the third kind, which I
shall divide by complete integrals of the first kind. For real $k^2<1$, let
\begin{equation}P(n,k)=\frac{\Pi(n,k)}{\Pi(0,k)},\quad 
\Pi(n,k)=\int_0^{\pi/2}\frac{{\rm d}\theta}{(1-n\sin^2\theta)\sqrt{1-k^2\sin^2\theta}}
\label{pink}\end{equation}
where $\Pi(0,k)=(\pi/2)/{\rm AGM}\,(1,\sqrt{1-k^2})$ is determined by the arithmetic-geometric mean of Gauss.
Then $P(n,k)$ is analytic in the $n$-plane with a cut $n\in[1,\infty]$ on which its principal value
is $1-P(k^2/n,k)$.

With $s=w^2$, integration over 3-body phase space~\cite{DD} determines
\begin{equation}k^2=1-\frac{16m_2m_3m_4w}{W},\quad W=(w_+^2-m_+^2)(w_-^2-m_-^2)\label{kp}\end{equation} 
with $w_\pm=w\pm m_2$ and $m_\pm=m_3\pm m_4$. Then I obtain
\begin{equation}\sigma^\prime_{2,3,4}(w^2)=
\frac{4\pi m_3m_4}{{\rm AGM}\,(\sqrt{16m_2m_3m_4w},\sqrt{W})}\Re\left(
\,\sum_{i=+,-}E_i\,\frac{P(n_i,k)-P(n_1,k)}{t_i-t_1}\right)\label{s234}\end{equation}
with coefficients and arguments given, as compactly as possible, by
\begin{gather}E_\pm=\frac{m_2^2-m_3^2+m_5^2}{2m_5^2}\pm
\left(\frac{m_4^2-m_5^2-w^2}{2m_5^2}\right)\,\frac{\Delta_{2,3,5}}{\Delta_{4,5}(w^2)},\label{ei}\\
t_\pm=\frac{\gamma\pm\Delta_{2,3,5}\Delta_{4,5}(w^2)}{2m_5^2},\quad t_1=m_1^2,\quad 
n_i=\frac{(w_-^2-m_+^2)(t_i-m_-^2)}{(w_-^2-m_-^2)(t_i-m_+^2)},\label{ti}\\
\gamma=(m_2^2+m_3^2+m_4^2-m_5^2+w^2)m_5^2+(m_2^2-m_3^2)(m_4^2-w^2).\label{ga}
\end{gather}
A strong check of~(\ref{s234}) is its hidden invariance under the non-trivial transformation
$(m_1,m_2,m_3,m_4,m_5)\to(m_5,m_4,m_3,m_2,m_1)$, despite division by
$m_5^2$ in~(\ref{ei},\ref{ti}). To obtain $\sigma^\prime_{1,3,5}$,
take the transform $(m_1,m_2,m_3,m_4,m_5)\to(m_4,m_5,m_3,m_1,m_2)$.

An AGM procedure evaluates $P(n,k)=\Pi(n,k)/\Pi(0,k)$ to high precision.
\begin{enumerate}
\item Initialize $[a,\,b,\,p,\,q]=[1,\,\sqrt{1-k^2},\,\sqrt{1-n},\,n/(2-2n)]$. Then set $f=1+q$.
\item Set $m=ab$ and then $r=p^2+m$. Compute a vector of new values as follows:
$[(a+b)/2,\,\sqrt{m},\,r/(2p),\,(r-2m)q/(2r)]$. Then replace $[a,\,b,\, p,\, q]$ by those new values. 
Then add $q$ to $f$.
\item If $|q/f|$ is sufficiently small, then return $P(n,k)=f$, else go to step 2.
\end{enumerate}
This converges very quickly, for $n\notin[1,\infty]$. For real $n\ge1$, invoke the procedure with $n$ replaced by
$k^2/n<1$, to obtain the principal value $\Re P(n,k)=1-P(k^2/n,k)$.

\subsection{Criterion for an anomalous contribution}

If there is an anomalous contribution, it occurs above the higher of the two-particle thresholds.
Without loss of generality, suppose that $s_{4,5}\ge s_{1,2}$. Then
\begin{equation}\sigma^\prime(s)=\sigma^\prime_{\rm N}(s)+\sigma^\prime_{\rm E}(s)+
C_{\rm A}\frac{\Theta(s-s_{4,5})}{\Delta_{4,5}(s)}\,\Re\left(\frac{2\pi{\rm i}\Delta_{4,5}(s_-)}{s-s_-}\right)
\label{ca}\end{equation}
with $C_{\rm A}\ne0$ if and only if $(m_1+m_2)(m_3^2+m_1m_2)<m_1m_5^2+m_2m_4^2$
and at least one of $\Delta_{1,3,4}$ and $\Delta_{2,3,5}$ is imaginary, in which case
$C_{\rm A}=\pm1$ is the sign of the imaginary part of $\Delta_{4,5}(s_-)$.
This value of $C_{\rm A}$ is required by the elliptic contribution at high energy. 
With $L_k=m_k^2\log(s/m_k^2)$, the large-$s$ behaviour
\begin{equation}s^2\sigma^\prime(s)=2L_3+\sum_{k=1,2,4,5}(L_k+m_k^2)
\,+\,O\left(\frac{\log(s)}{s}\right)\label{asy}\end{equation}
invariably holds. The elliptic contribution $\sigma^\prime_{\rm E}$ 
in~(\ref{ca}) is oblivious to the anomalous threshold problem. Its high-energy behaviour
determines $C_{\rm A}$, ensuring~(\ref{asy}).
Thus the leading singularity of Lev Landau in the anomalous non-elliptic sector
neatly performs a task set by Carl Friedrich Gauss in the elliptic sector.
I tested this at 50-digit precision with more than a million random assignments of masses.
Each test took less than a millisecond, thanks to the extraordinary efficiency of the AGM.

\section{Zero-mass limits}

Despite division by $m_3^2$ in~(\ref{albe},\ref{spm}), the two-particle
cuts give finite contributions as $m_3\to0$, except in the special case that $m_1=m_4$ and $m_2=m_5$.
As $m_3\to0$ with $m_1\ne m_4$ and $m_2\ne m_5$, diligent algebra shows that
\begin{gather}
\Delta_{1,2}(s)\sigma_{1,2}^\prime(s)\to\Re\left((2s-s_3)D_{4,5}(s)
+\widehat{L}_{4,5}+\sum_{i=0,3}C_i\,\frac{D_{4,5}(s)-D_{4,5}(s_i)}{s-s_i}\right),\label{s120}\\
s_3=\frac{(m_2^2m_4^2-m_1^2m_5^2)(m_1^2-m_2^2-m_4^2+m_5^2)}{M},\,
M=(m_1^2-m_4^2)(m_2^2-m_5^2),\label{s3}\\
\widehat{L}_{4,5}=\log\left(\frac{m_4^2m_5^2}{M}\right),\,
C_3=-\left(\frac{m_1^2}{u}-m_2^2u\right)\left(\frac{m_4^2}{u}-m_5^2u\right),\,
u=\frac{m_1^2-m_4^2}{m_2^2-m_5^2}\label{c3}
\end{gather}
with $\sigma_{4,5}^\prime$ given by $(m_1,m_2)\leftrightarrow(m_4,m_5)$. 
There is no anomalous term, since both $\Delta_{1,3,4}$ and $\Delta_{2,3,5}$ are real for $m_3=0$.
As $m_3\to0$ with $m_1=m_4$
\begin{align}
\Delta_{2,4}(s)\sigma_{1,2}^\prime(s)&\to\Re\left((3s-m_2^2-2m_4^2-m_5^2)D_{4,5}(s)
+\log\left(\frac{m_4m_5^3}{(m_2^2-m_5^2)^2}\right)\right.\nonumber\\
&+\left.(m_2^2-m_4^2)(m_4^2-m_5^2)\,\frac{D_{4,5}(s)-D_{4,5}(0)}{s}\right)\label{s12e}
\end{align}
with $\sigma_{4,5}^\prime$ given by $m_2\leftrightarrow m_5$.
The degenerate case with $m_2=m_5$ will be considered after
adding contributions from three-particle cuts.

At $m_3=0$, the three-particle cuts yield logarithms. For $w>m_j+m_k$, let
\begin{equation}\widehat{P}_{j,k}(t,w)=\frac{(m_k^2-t)v(t)}{(w-m_j)^2-t}
\log\left(\frac{v(t)+v(m_k^2)}{v(t)-v(m_k^2)}\right),\; 
v(t)=\left(\frac{(w-m_j)^2-t}{(w+m_j)^2-t}\right)^{1/2}\label{pjk}\end{equation}
where $v(m_k^2)$ is real and positive, while $v(t)$ may be complex. As $m_3\to0$
\begin{equation}\sigma_{2,3,4}^\prime(w^2)\to\Re\left(\sum_{i=\pm}E_i\,
\frac{\widehat{P}_{2,4}(t_i,w)-\widehat{P}_{2,4}(m_1^2,w)}{t_i-m_1^2}\right)
\label{s204}\end{equation}
with coefficients and arguments in~(\ref{ei},\ref{ti}) that are well behaved at $m_3=0$.

With $m_1=m_4$ and $m_2=m_5$ all thresholds collide as $m_3\to0$, giving~\cite{B90}
\begin{gather}
\sigma^\prime(s)\to\Theta(s-s_{4,5})\,\frac{2\mu(y_4)+2\mu(y_5)-8\mu(y_4y_5)}{\Delta_{4,5}(s)},\label{m45}\\
\mu(y)=\log|1-y|+\frac{y\log|y|}{1-y},\quad y_k=\frac{-2m_k^2}{s-m_4^2-m_5^2+\Delta_{4,5}(s)}.\label{muy} 
\end{gather}

Next, consider cases with $m_3>0$ and one of the other masses vanishing.
As $m_4\to0$, logarithms from~(\ref{pjk}) appear in
\begin{equation}\sigma_{2,3,4}^\prime(w^2)\to\Re\left(\sum_{i=\pm}E_i\,
\frac{\widehat{P}_{2,3}(t_i,w)-\widehat{P}_{2,3}(m_1^2,w)}{t_i-m_1^2}\right).\label{p23}\end{equation}
The logarithms in~(\ref{s12}) are modified, to give
\begin{gather}\Delta_{1,2}(s)\sigma_{1,2}^\prime(s)\to\Re\left((s+\alpha)\widehat{D}_5(s)
+\widehat{L}_5+\sum_{i=0,+,-}C_i\,\frac{\widehat{D}_{5}(s)-\widehat{D}_{5}(s_i)}{s-s_i}\right),\label{d5}\\
\widehat{D}_5(s)=\frac{1}{s-m_5^2}\log\left(1-\frac{s}{m_5^2}\right),\quad 
\widehat{L}_5=\log\left(\frac{m_5^2}{m_3^2}\right).\label{ld5}\end{gather}

An elliptic contribution persists if two non-adjacent edges have vanishing mass. As $m_1\to0$ and $m_5\to0$
\begin{equation}(w^2-m_4^2)\sigma^\prime_{2,3,4}(w^2)\,\to\,
-\frac{4\pi m_3m_4\Re R(w^2,m_2^2,m_3^2,m_4^2)}{{\rm AGM}\,(\sqrt{16m_2m_3m_4w},\sqrt{W})}\end{equation}
where $R$ contains, in general, three elliptic terms. When $m_2\ne m_3$, I obtain
\begin{gather}R(s,b,c,d)=P(\widehat{n},k)-\rho P(n_0,k)+(\rho-1)P(n_3,k),\\
\widehat{n}=\frac{w_-^2-m_+^2}{w_-^2-m_-^2},\, \frac{n_0}{\widehat{n}}=\frac{m_-^2}{m_+^2},
\, \frac{n_3}{\widehat{n}}=\frac{t_3-m_-^2}{t_3-m_+^2},\,
t_3=\frac{(bd-cs)(b-c+d-s)}{(b-c)(d-s)},\label{R}\\
\rho=\left(\frac{d-s}{b-c+d-s}\right)\left(\frac{(b+c)(d-s)+(b-c)(b+d)}{bd-cs}\right)\label{rho}\end{gather}
with $(s,b,c,d)=(w^2,m_2^2,m_3^2,m_4^2)$.
In equal-mass cases this simplifies, giving
\begin{equation}R(s,c,c,d)=2P(\widehat{n},k)-2P(n_0,k),\quad 
R(s,d,d,d)=\frac{s-9d}{6d}\label{req}\end{equation}
with a notable rational result for $m_2=m_3=m_4$, proven in~\cite{B90}.

Remaining cases with vanishing masses are logarithmic and hence somewhat easier.
Care is needed in cases like~(\ref{m45}) when thresholds collide. In the bivariate case 
where $m_1$, $m_3$ and $m_4$ vanish, with $m_2\ne m_5$, I obtain
\begin{gather}s\sigma^\prime(s)\to
\Theta(s-m_2^2)\,\tau(x_2,x_5)+\Theta(s-m_5^2)\,\tau(x_5,x_2),\quad x_k=\frac{m_k^2}{s},\label{w25}\\
\quad\tau(x,y)=\nu(x,y)-\nu(y,x)-\mu(y),\quad\nu(x,y)=\frac{2}{1-x}\log\left|\frac{1-y}{x-y}\right|\quad
\end{gather}
after taking account of the threshold collisions $s_{2,3,4}\to s_{1,2}$ and $s_{1,3,5}\to s_{4,5}$.

\section{Tadpoles and number theory}

Since $s\sigma^\prime(s)$ is dimensionless it depends only on the 5 ratios $m_k^2/s$, with $k\in[1,5]$.
Tetrahedral tadpoles have a simple behaviour under the rescaling $m_k\to\lambda\,m_k$, with $k\in[1,6]$.
This gives $F\to F+12\zeta_3\log(\lambda)$ for the finite part $F$ in~(\ref{f123}).
To standardize, I set both $\overline{m}$ and the largest $m_k$ to unity.

\subsection{Perfect tadpoles}

I define a tetrahedral tadpole to be {\em perfect} if and only if the
K\"all\'en function vanishes at each of its 4 vertices, thereby avoiding all resolutions
of square roots. I promote the subscripts and superscripts of $F$
to arguments that denote the 6 masses and define a two-parameter family of perfect tadpoles
\begin{equation}\widehat{F}(x,y)=F_{(x,y,1)}^{(1-y,1-x,|x-y|)}=\widehat{F}(y,x)=\widehat{F}(1-x,1-y)\label{fh}\end{equation}
with symmetries restricting distinct cases to $x\ge y\ge1-x\ge0$ and hence $x\in[\frac12,1]$.
In~\cite{B99}, I identified the tetralogarithms in two perfect tadpoles,
\begin{gather}
\widehat{F}(1,0)=F_{(1,0,1)}^{(1,0,1)}=17\zeta_4+16U_{3,1},\quad
\widehat{F}(1,1)=F_{(1,1,1)}^{(0,0,0)}=12\zeta_4\label{fun},\\
U_{3,1}=\sum_{m>n>0}\frac{(-1)^{m+n}}{m^3n}=
\tfrac12\zeta_4+\tfrac{1}{2}\zeta_2\log^2(2)
-\tfrac{1}{12}\log^4(2)-2\,{\rm Li}_4(\tfrac12).\label{u31}
\end{gather}

\subsection{Fast elliptic determination of a perfect tadpole}

Now consider an elliptic route to evaluating $\widehat{F}(\frac12,\frac12)$.
With $(m_3,m_6)=(1,0)$ and $m_1=m_2=m_4=m_5=\frac12$, I obtained
\begin{gather}\widehat{F}(\tfrac12,\tfrac12)
=\tfrac12\int_1^\infty{\rm d}s\left(\widehat{\sigma}^\prime_{\rm N}(s)+\widehat{\sigma}^\prime_{\rm E}(s)\right)\log^2(s),\label{sne}\\
w^2\widehat{\sigma}^\prime_{\rm N}(w^2)=\Theta(w-1)\left(2\log\left(\frac{r+1}{r-1}\right)-4r\log(2)\right),\;
r=\frac{w}{\sqrt{w^2-1}},\label{snp}\\ 
w^2\widehat{\sigma}^\prime_{\rm E}(w^2)=
\frac{4\pi(1-P(n,k))\Theta(w-2)}{{\rm AGM}(2\sqrt{w},(w-1)\sqrt{w^2+2w})},\;
n=\frac{w^2-2w}{(w-1)^2},\;\frac{k^2}{n}=\frac{(w+1)^2}{w^2+2w}\label{sep}\end{gather}
and readily discovered a new reduction of a perfect tadpole to tetralogarithms 
\begin{equation}\widehat{F}(\tfrac12,\tfrac12)=F_{(\frac12,\frac12,1)}^{(\frac12,\frac12,0)}
=30\,\zeta_3\log(2)-16\,\zeta_4-32\,U_{3,1}.\label{fhh}\end{equation}

\subsection{Relations between tadpoles}

\mbox{\hspace{1cm}}\hfill
\dia{\tetmrk{\hf}{\hf}1{\hf}{\hf}0}{}
\dia{\tetmrk1{\hf}01{\hf}{\hf}}{}
\dia{\tetmrk{\hf}{\hf}1{\hf}{\hf}1}{}
\mbox{\hspace{1cm}}
\begin{center}{\bf Figure 2:} The perfect tadpoles $\widehat{F}(\frac12,\frac12)$,
$\widehat{F}(1,\frac12)$ and $\widehat{G}(\frac12)$ in relation~(\ref{rel})\end{center}

In addition to the two-parameter family $\widehat{F}(x,y)$ in~(\ref{fh})
there is a one-parameter family $\widehat{G}(x)=F_{(x,1-x,1)}^{(x,1-x,1)}$ of perfect tadpoles,
with $x\in[\frac12,1]$ and $\widehat{G}(1)=17\zeta_4+16U_{3,1}$.

I used the efficient AGM of Gauss to obtain 200 good digits of
\begin{equation}\widehat{G}(\tfrac12)=F_{(\frac12,\frac12,1)}^{(\frac12,\frac12,1)}=
-\int_1^\infty{\rm d}s\left(\widehat{\sigma}^\prime_{\rm N}(s)
+\widehat{\sigma}^\prime_{\rm E}(s)\right)
{\rm Li}_{2}(1-s)\label{gh}\end{equation}
to which all routes are elliptic. This revealed the intriguing empirical relation 
\begin{equation}
2\widehat{F}(\tfrac12,\tfrac12)+2\widehat{F}(1,\tfrac12)+\widehat{G}(\tfrac12)=42\,\zeta_4+24\,\zeta_3\log(2)
\label{rel}\end{equation}
with a non-elliptic route to $\widehat{F}(1,\tfrac12)$
that leads to multiple polylogarithms in an alphabet of differential forms, 
${\rm d}x/(x-a_i)$, with $a_i\in\{0,1,-1,-2\}$. With $L=\log(2)$, relation~(\ref{rel}) gives an empirical reduction
to classical polylogs in
\begin{equation}
\widehat{G}(\tfrac12)=6\left(2\zeta_4-3{\rm Li}_4(\tfrac14)\right)
+8\left(2\zeta_3-3{\rm Li}_3(\tfrac14)\right)L
-12\,{\rm Li}_2(\tfrac14)L^2-4L^4.\label{ghis}\end{equation}

There are linear relations between binary tadpoles, as here
\begin{align}
3F_{(0,0,0)}^{(1,1,1)}&=F_{(1,1,1)}^{(0,0,0)}+2F_{(1,1,0)}^{(1,0,0)}\label{v3l}\\
3F_{(1,1,0)}^{(0,0,0)}&=F_{(1,0,0)}^{(0,0,0)}+2F_{(1,1,1)}^{(0,0,0)}\label{v2a}\\
F_{(1,1,1)}^{(1,1,1)}+F_{(1,0,0)}^{(1,0,0)}&=F_{(1,0,1)}^{(1,0,1)}+F_{(0,0,0)}^{(1,1,1)}\label{v2m}
\end{align}
with tetralogarithms of second and sixth roots of unity neatly cancelling~\cite{B99}.

Relation~(\ref{v2m}) is empirical. Like Figure~2, it relates a completely massive elliptic tadpole
to cases with zero mass that have non-elliptic routes. In the course of the current work, I checked~(\ref{v2m})
at 200-digit precision.\\
\mbox{\hspace{1cm}}\hfill
\dia{\tetmrk011010}{}
\dia{\tetmrk100101}{}
\dia{\tetmrk110010}{}
\mbox{\hspace{1cm}}
\begin{center}{\bf Figure 3:} Binary tadpoles in relation~(\ref{v3l})\end{center}

\subsection{Number fields of the alphabets of tadpoles}

So far, one might guess that a tadpole with rational masses evaluates to multiple tetralogarithms
in an alphabet whose number field is no larger than the compositum 
of the quadratic number fields associated by K\"all\'en to the vertices of the tetrahedron,
namely the field $Q(\Delta_{1,3,4},\Delta_{2,3,5},\Delta_{1,2,6},\Delta_{4,5,6})$. Yet that is {\em not} the case.

Binary tadpoles evaluate to multiple polylogarithms in
an alphabet containing sixth roots of unity, with $\lambda=(1+\sqrt{-3})/2$
appearing if three massive edges meet at a vertex, where $\Delta_{i,j,k}=\sqrt{-3}$.
With ${\rm Cl}_2(\pi/3)=\Im{\rm Li}_2(\lambda)$, the imperfect binary tadpoles with 5 and 6 unit masses give
\begin{align}
F_{(1,1,1)}^{(1,1,0)}&=\tfrac{550}{27}\,\zeta_4-\tfrac83\,{\rm Cl}_2^2(\pi/3)
+16\sum_{m>n>0}\frac{(-1)^m\cos(2\pi n/3)}{m^3n},\label{F5}\\
F_{(1,1,1)}^{(1,1,1)}&=16\,\zeta_4+4\,{\rm Cl}_2^2(\pi/3)+8\,\sum_{m>n>0}\frac{(-1)^{m+n}}{m^3n}.\label{F6}
\end{align}
A counterexample to the idea that the K\"all\'en number field
is large enough to furnish the alphabet of a specific tadpole is provided by relation~(\ref{v3l}) between the tadpoles of
Figure~3. The first contains ${\rm Cl}_2^2(\pi/3)$. The second is perfect and evaluates to $12\zeta_4$. So far, so good.
The problem is with the third tadpole. While imperfect, it has a K\"all\'en field that is rational, with
$\Delta_{i,j,k}\in\{0,1\}$. Yet its alphabet must involve $\lambda=\exp(2\pi{\rm i}/3)$, because of relation~(\ref{v3l}).

Faced with this limited, yet potent, evidence, I arrive at three {\em suggestions}, each too weak to be dignified as a well-tested conjecture.
\begin{enumerate}
\item Tetrahedral tadpoles with rational masses may reduce to multiple or single tetralogarithms
whose alphabet lies in an algebraic number field.
\item If the tadpole is perfect, the alphabet may be rational.
\item The alphabetic field of an imperfect tadpole may include the K\"all\'en field.
\end{enumerate}

As an {\em experimentum crucis},  I evaluated another totally massive imperfect tadpole
with K\"all\'en field $Q(\sqrt{-3})$. Seeking a reduction to tetralogarithms in an alphabet
$\{0,1,-1,-2,\lambda\}$, I found an empirical reduction to polylogs,
\begin{equation}F_{(\frac12,\frac12,\frac12)}^{(1,1,1)}=3\,\zeta_3\log(2)-4\,U_{3,1}+10\left({\rm Cl}_2^2(\pi/3)+\zeta_4\right)
-\tfrac12\widehat{G}(\tfrac12).\label{fhhh}\end{equation}

\section{Comments and summary}

Generic 5-mass kites and 6-mass tadpoles present few 
conceptual problems: 54 years ago, Don Perkins taught me that experimental data on the then topical decay
$\eta\to3\pi$ is best displayed on a Dalitz plot~\cite{DP} whose area is an elliptic integral. Later that
year, Gabriel Barton taught me that this elliptic integral gives the discontinuity of a two-loop
two-point function with three internal massive particles. It was not hard to see how to extend
that to the kite with 5 arbitrary masses. Diligent algebra and use of Cauchy's integral theorem leads to the generic 
formul{\ae} of Section~2, which I strove to reduce to their essentials, in a compact
manner that leads to efficient coding, thanks to the AGM of Gauss for complete
elliptic integrals, whatever their kind.

There remained a problem of principle. The kite has no anomalous thresholds, yet
the triangles that it contains may have dispersion relations with anomalous thresholds~\cite{AF}
and complex branch points. Here I was
both challenged and encouraged by Spencer Bloch, who knew from algebraic
geometry that I was skating on thin ice, by taking principal values to obtain
a discontinuity as the real part of a combination of multi-sheeted complex functions. 
Memories of Stanley Mandelstam's ominous drawing in~\cite{SM}, depicting contour deformation,
came to haunt me. Yet I was protected by Gabriel Barton, who had provided the key, in 1968.
 \begin{enumerate}
\item If you know the discontinuity $\sigma$ of $f$, use a dispersion relation to get $f$.
\item If you know only $\sigma^\prime$, integrate it against a log.
\item For the 4th-order photon propagator, $\sigma^\prime$ has logs, so $f$ has Lewin's tri-logs.
\item For the 4th-order electron propagator, $\sigma^\prime$ is elliptic, so $f$ is harder to get.
\item K\"all\'en did the easier case~\cite{KS}, leaving the harder case~\cite{S} for Sabry. 
\end{enumerate}

Working with $\sigma^\prime$,
I knew that any anomalous term
would be algebraic, not an integral. Agreement with Gauss,
at high energy, gave the rule for the anomalous coefficient $C_{\rm A}$
in~(\ref{ca}), without having to deform any contour. 

Turning to specifics, I remark how much more work one needs to do when masses
tend towards zero and thresholds collide. The algebra in Section~3 was demanding and was enabled
by efficient partial fractioning and multivariate factorization in Tony Hearn's open-source {\sc Reduce}, 
which is freely available~\cite{AH}. The rest of the work was done in {\sc Pari/GP}, which is also freely
available~\cite{GP}. There was no need for expensive proprietary software. 
A modest laptop delivers 200 digits of a kite or tadpole in a few seconds.

\subsection{Summary}

Elliptic substructure of 2-loop kites and 3-loop tadpoles is not an obstacle to fast numerical evaluation.
The time taken to evaluate a complete elliptic integral, of whatever kind,
is commensurate with that for a logarithm and less than that for a dilogarithm.
Thanks to Gauss, elliptic integrals should be embraced, not feared.

Anomalous terms are not a problem. They submit to Gauss, at high energy.

Benchmarks for kites, given at low precision in~\cite{BB,M}, were confirmed and easily extended to 600-digit precision.

The number theory of tadpoles is subtle. They may be polylogarithmic,
even in totally massive cases such as those in~(\ref{ghis},\ref{F6},\ref{fhhh}), to which every route has an elliptic obstruction.

Subsection~4.4 gives far-reaching suggestions on the number theory of tadpoles. They set demanding
challenges for adept users
of packages such as Erik Panzer's {\sc HyperInt}~\cite{HI,HI1,YZ5} and
{\sc MZIteratedIntegral}~\cite{KCA} from Kam Cheong Au,
to investigate these suggestions analytically.

\newpage

\begin{acknowledgement} 
First and foremost, I thank the departed dead. 
The influences of Gabriel Barton (1934--2022) and Donald Hill Perkins (1925--2022) 
have already been made clear. I thank the brothers
Borwein, Jonathan (1951--2016) and Peter (1953--2020), for
their fine monograph on the AGM, and Freeman Dyson (1924--2020)
for encouraging my efforts to
relate quantum field theory to number theory. Turning to the
living, I thank Dirk Kreimer and his entire Arbeitsgruppe,
who made my many visits to Berlin so enjoyable and
profitable. I thank mathematical advisors for their patient
instruction. In particular,
the influence of Spencer Bloch, Francis Brown, Pierre
Deligne and Don Zagier extends deeply into the work reported
here, in both the polylogarithmic and elliptic sectors. I am grateful to
generous hosts for their invitations to
workshops and conferences at the many venues mentioned in the introduction.
I thank Steven Charlton and Yajun Zhou, for advice on polylogarithms,
Kevin Acres and Michael St Clair Oakes, for close scrutiny
of a draft version, and William Edward Makin, for more than 60 years of 
discussion of natural philosophy and its firm relation
to other human endeavours.
\end{acknowledgement}


\begin{thebibliography}{99.}

\bibitem{Sm1}
Luise Adams, Christian Bogner and Stefan Weinzierl, 
{\em The two-loop sunrise graph in two space-time dimensions with arbitrary masses in terms of elliptic dilogarithms},
J.\ Math.\ Phys.\ {\bf 55} (2014) 102301.

\bibitem{AF} 
D.\ Amati and S.\ Fubini,
{\em Dispersion relation methods in strong interactions},
Annual Rev.\ Nucl.\ Sci.\ {\bf 12} (1962) 359-434,\\
\url{http://doi.org/10.1146/annurev.ns.12.120162.002043}~.

\bibitem{KCA}
Kam Cheong Au, 
{\em Iterated integrals and multiple polylogarithms at algebraic arguments},
arXiv:2201.01676 [math.NT].

\bibitem{BBBG}
D.H.\ Bailey, J.M.\ Borwein, D.\ Broadhurst and M.L.\ Glasser,
{\em Elliptic integral evaluations of Bessel moments},
J.\ Phys.\ A {\bf 41} (2008) 205203.

\bibitem{GB} 
G.\ Barton,
{\em Introduction to dispersion techniques in field theory}
(Benjamin, New York, 1965).

\bibitem{BB}
S.\ Bauberger and M.\ B\"ohm,
{\em Simple one-dimensional integral representations for two-loop self-energies: the master diagram},
Nucl.\ Phys.\ B {\bf 445} (1995) 25-46.

\bibitem{BV}
Spencer Bloch and Pierre Vanhove,
{\em The elliptic dilogarithm for the sunset graph},
J.\ Number Theory {\bf 148} (2015) 328-364.

\bibitem{BKV}
Spencer Bloch, Matt Kerr and Pierre Vanhove,
{\em A Feynman integral via higher normal functions},
Compositio Math.\ {\bf 151} (2015) 2329-2375.

\bibitem{Dev4}
Johannes Bl\"umlein,
{\em Iterative non-iterative integrals in quantum field theory},
in Texts and Monographs in Symbolic Computation, 
{\em Elliptic Integrals, elliptic functions and modular forms in quantum field theory}
(Springer, 2019).

\bibitem{MPL4}
J.\ Bl\"umlein,  D.\ Broadhurst and J.A.M.\ Vermaseren,
{\em The multiple zeta value data mine},
Comput.\ Phys.\ Commun.\ {\bf }181 (2010) 582-625.

\bibitem{AGM1}
Jonathan and Peter Borwein, {\em Pi and the AGM} (Wiley, NJ, 1987).

\bibitem{MPL2}
J.M.\ Borwein, D.M.\ Bradley, D.\ Broadhurst and P.\ Lisonek,
{\em Special values of multiple polylogarithms},
Trans.\ Am.\ Math.\ Soc.\ {\bf 353} (2001) 907-941.

\bibitem{SYM2}
Jacob L.\ Bourjaily, Andrew J.\ McLeod, Marcus Spradlin, Matt von Hippel and Matthias Wilhelm,
{\em The elliptic double-box integral: massless amplitudes beyond polylogarithms},
Phys.\ Rev.\ Lett.\ {\bf 120} (2018) 121603.

\bibitem{AGM2}
Richard P.\ Brent,
{\em The Borwein brothers, Pi and the AGM},
Springer Proc.\  Math.\ Stat.\ {\bf 313} (2020) 323-348,
arXiv:1802.07558 [math.NT].
 
\bibitem{B90}
D.\ Broadhurst, {\em The master two loop diagram with masses},
Z.\ Phys.\ C {\bf 47} (1990) 115-124.

\bibitem{B92}
D.\ Broadhurst,
{\em Three loop on-shell charge renormalization without integration: $\Lambda^{\overline{MS}}_{QED}$ to four loops},
Z.\ Phys.\ C {\bf 54} (1992) 599-606.

\bibitem{B99}
D.\ Broadhurst,
{\em Massive 3-loop Feynman diagrams reducible to SC* primitives of algebras of the sixth root of unity},
Eur.\ Phys.\ J.\ C {\bf 8} (1999) 311-333.

\bibitem{Bc3}
D.\ Broadhurst,
{\em Feynman integrals, L-series and Kloosterman moments},
Commun.\ Number Theory Phys.\ {\bf 10} (2016) 527-569.

\bibitem{Bc2}
D.\ Broadhurst and A.\ Mellit,
{\em Perturbative quantum field theory informs algebraic geometry},
PoS LL2016 (2016) 079.

\bibitem {Bc4} D.\ Broadhurst and D.P.\ Roberts,
{\em Quadratic relations between Feynman integrals},
PoS LL2018 (2018) 053.

\bibitem{Bc1}
D.\ Broadhurst and O.\ Schnetz,
{\em Algebraic geometry informs perturbative quantum field theory},
PoS LL2014 (2014) 078.

\bibitem{SR1}
D.\ Broadhurst, J.\ Fleischer and O.V.\ Tarasov,
{\em Two-loop two-point functions with masses: asymptotic expansions and Taylor series in any dimension},
Z.\ Phys.\ C {\bf 60} (1993) 287-302.

\bibitem{B93}
D.\ Broadhurst, A.L.\ Kataev and O.V.\ Tarasov,
{\em Analytical on-shell QED results: 3-loop vacuum polarization, 
4-loop beta-function and the muon anomaly},
Phys.\ Lett.\ B {\bf 298} (1993) 445-452.

\bibitem{Dev3}
Johannes Broedel, Claude Duhr, Falko Dulat and Lorenzo Tancredi,
{\em Elliptic polylogarithms and iterated integrals on elliptic curves II: an application to the sunrise integral},
Phys.\ Rev.\ D {\bf 97} (2018) 116009.

\bibitem{St2}
Johannes Broedel, Oliver Schlotterer and Federico Zerbini, 
{\em From elliptic multiple zeta values to modular graph functions: open and closed strings at one loop},
J.\ High Energy Phys.\ {\bf  2019} (2019) 155.

\bibitem{Dev1} 
Francis Brown and Andrey Levin,
{\em Multiple elliptic polylogarithms}, \\
arXiv:1110.6917 [math.NT].

\bibitem{SYM1}
Simon Caron-Huot and Kasper J.\ Larsen,
{\em Uniqueness of two-loop master contours},
J.\ High Energy Phys.\ {\bf 2012} (2012) 26.

\bibitem{SC} 
S.\ Charlton, H.\ Gangl, D.\ Radchenko and D.\ Rudenko,
{\em On the Goncharov depth conjecture and polylogarithms of depth two},
arXiv:2210.11938 [math.NT].

\bibitem{DP} 
A.M.\ Cnops, G.\ Finocchiaro, P.\ Mittner, J.P.\ Dufey, B.\ Gobbi, M.A.\ Pouchon and A.\ M\"uller,
{\em Dalitz plot analysis of the decay $\eta\to\pi^+\pi^-\pi^0$},
Phys.\ Lett.\ B {\bf 27} (1968) 113-116.

\bibitem{DD}
A.I.\ Davydychev and R.\ Delbourgo,
{\em Explicitly symmetrical treatment of three-body phase space},
J.\ Phys.\ A {\bf 37} (2004) 4871.

\bibitem{MPL5}
Pierre Deligne,
{\em Le groupe fondamental motivique de ${\bf G}_m-\mu_N$ pour $N=2$,3,4,6 ou 8},
Pub.\ Math.\ IHES {\bf 112} (2010) 101-141.

\bibitem{St1}
Eric D'Hoker, Michael B.\ Green and Pierre Vanhove,
{\em Proof of a modular relation between 1-, 2- and 3-loop Feynman diagrams on a torus},
J.\ Number Theory {\bf 196} (2019) 381-419.

\bibitem{MG} M.L.\ Goldberger,
{\em Causality conditions and dispersion relations},
Phys.\ Rev.\ {\bf 99} (1955) 979-985.

\bibitem{AH} 
Anthony C.\ Hearn,
{\sc Reduce}, {\em a portable general-purpose computer algebra system},
\url{http://reduce-algebra.sourceforge.io/about.php} (2022).

\bibitem{MPL3}
Kentaro Ihara, Masanobu Kaneko and Don Zagier,
{\em Derivation and double shuffle relations for multiple zeta values},
Compositio Math.\ {\bf 142} (2006) 307-338.

\bibitem{GJ}
G.S.\ Joyce, {\em On the simple cubic lattice Green function}, 
Phil.\ Trans.\ Roy.\ Soc.\ {\bf 273} (1973) 583-610.

\bibitem{Kb}
G.\ K\"all\'en, 
{\em Elementary Particle Physics}
(Addison-Wesley, Mass., 1964).

\bibitem{KS}
G.\ K\"all\'en and A.\ Sabry, 
{\em Fourth order vacuum polarization},
Math.\ Fys.\ Medd.\ Dan.\ Vid Selsk {\bf 29} (1955) 17,\\
\url{http://inspirehep.net/files/6bb500b7fc71eced402f9824fbe96f65}~.

\bibitem{Sm2}
Dirk Kreimer,
{\em Bananas: multi-edge graphs and their Feynman integrals},
Lett.\ Math.\ Phys.\ {\bf 113} (2023) 28.

\bibitem{KK}
R.\ de L.\ Kronig and H.A.\ Kramers, 
{\em  Zur Theorie der Absorption und Dispersion in den R\"ontgenspektren},
Z.\ Phys.\ {\bf 48} (1928) 174-179.

\bibitem{Sm3}
Pierre Lairez and Pierre Vanhove,
{\em Algorithms for minimal Picard-Fuchs operators of Feynman integrals},
Lett.\ Math.\ Phys.\ {\bf 113} (2023) 37.

\bibitem{SR2}
Stefano Laporta,
{\em Analytical expressions of three- and four-loop sunrise Feynman integrals and
four-dimensional lattice integrals},
Int.\ J.\ Mod.\ Phys.\ A  {\bf 23} (2008) 5007-5020.

\bibitem{SL}
Stefano Laporta,
{\em High-precision calculation of the 4-loop contribution to the electron $g-2$ in QED},
Phys.\ Lett.\ B {\bf 772} (2017) 232-238.

\bibitem{L}
 L.\ Lewin, 
{\em Dilogarithms and associated functions} (Macdonald, London, 1958);
{\em Polylogarithms and associated functions} (North-Holland, Amsterdam, 1981).

\bibitem{SM}
S.\ Mandelstam,
{\em Unitarity condition below physical thresholds in the normal and anomalous cases},
Phys.\ Rev.\ Lett.\ {\bf 4} (1960) 84-87.

\bibitem{M}
Stephen P.\ Martin,
{\em Evaluation of two-loop self-energy basis integrals using differential equations},
Phys.\ Rev.\ D {\bf 68} (2003) 075002.

\bibitem{HI}
Erik Panzer,
{\em Algorithms for the symbolic integration of hyperlogarithms with applications to Feynman integrals},
Computer Phys.\ Comm.\ {\bf 188} (2015) 148-166.

\bibitem{HI1} 
Erik Panzer, 
{\sc HyperInt},\\
\url{http://bitbucket.org/PanzerErik/hyperint/src/master/} (2023).

\bibitem{GP}
PARI Group, 
{\sc Pari/GP} version 2.15.4, Univ.\ Bordeaux,\\
\url{http://pari.math.u-bordeaux.fr/} (2023).

\bibitem{Dev5}
Sebastian P\"ogel, Xing Wang and Stefan Weinzierl, 
{\em The three-loop equal-mass banana integral in $\epsilon$-factorised form with meromorphic modular forms},
J.\ High Energy Phys.\ {\bf 2022} (2022) 62.

\bibitem{Dev2}
E.\ Remiddi and L.\ Tancredi,
{\em An elliptic generalization of multiple polylogarithms},
Nucl.\ Phys.\ B {\bf 925} (2017) 212-251.

\bibitem{MPL1} 
E.\ Remiddi and J.A.M.\ Vermaseren,
{\em Harmonic polylogarithms},
Int.\ J.\ Mod.\ Phys.\ A {\bf 15} (2000) 725-754.

\bibitem{S} A.\ Sabry,
 {\em Fourth order spectral functions for the electron propagator}, 
 Nucl.\ Phys.\ {\bf 33} (1962) 401-430.

\bibitem{MS} M.\ Steinhauser,
{\em MATAD: a program package for the computation of MAssive TADpoles},
Computer Phys.\ Comm.\ {\bf 134} (2001) 335-364.

\bibitem{St3}
Don Zagier and Federico Zerbini,
{\em Genus-zero and genus-one string amplitudes and special multiple zeta values},
Commun.\ Number Theory Phys.\  {\bf 14} (2020) 413-452.

\bibitem{YZ1}
Yajun Zhou,
{\em Wick rotations, Eichler integrals and multi-loop Feynman diagrams},
Commun.\ Number Theory Phys.\ {\bf 12} (2018) 127-192.

\bibitem{YZ2}
Yajun Zhou, 
{\em Hilbert transforms and sum rules of Bessel moments},
Ramanujan J.\ {\bf 48} (2019) 159-172.

\bibitem{YZ3}
Yajun Zhou,
{\em On Laporta’s 4-loop sunrise formul{\ae}}.
Ramanujan J.\ {\bf 50} (2019) 465-503.

\bibitem{YZ4}
Yajun Zhou,
{\em Wronskian algebra and Broadhurst-Roberts quadratic relations},
Commun.\ Number Theory Phys.\ {\bf 15} (2021) 651-741.

\bibitem{YZ5}
Yajun Zhou,
{\em Hyper-Mahler measures via Goncharov-Deligne cyclotomy},\\
arXiv:2210.17243 [math.NT] .

\end{thebibliography}
\end{document}